# The MIT Supercloud Dataset

Siddharth Samsi[*][§], Matthew L Weiss[*], David Bestor[*], Baolin Li[‡], Michael Jones[*], Albert Reuther[*],
Daniel Edelman[*], William Arcand[*], Chansup Byun[*], John Holodnack[*], Matthew Hubbell[*],
Jeremy Kepner[*], Anna Klein[*], Joseph McDonald[*], Adam Michaleas[*], Peter Michaleas[*],
Lauren Milechin[*], Julia Mullen[*], Charles Yee[*], Benjamin Price[*], Andrew Prout[*], Antonio Rosa[*],
Allan Vanterpool[†], Lindsey McEvoy[†], Anson Cheng[†], Devesh Tiwari[‡], Vijay Gadepally[*]
[*] MIT, [‡] Northeastern University, [†] US Air Force

*Abstract*—Artificial intelligence (AI) and Machine learning (ML) workloads are an increasingly larger share of the compute workloads in traditional High-Performance Computing (HPC) centers and commercial cloud systems. This has led to changes in deployment approaches of HPC clusters and the commercial cloud, as well as a new focus on approaches to optimized resource usage, allocations and deployment of new AI frameworks, and capabilities such as Jupyter notebooks to enable rapid prototyping and deployment. With these changes, there is a need to better understand cluster/datacenter operations with the goal of developing improved scheduling policies, identifying inefficiencies in resource utilization, energy/power consumption, failure prediction, and identifying policy violations. In this paper we introduce the MIT Supercloud Dataset which aims to foster innovative AI/ML approaches to the analysis of large scale HPC and datacenter/cloud operations. We provide detailed monitoring logs from the MIT Supercloud system, which include CPU and GPU usage by jobs, memory usage, file system logs, and physical monitoring data. This paper discusses the details of the dataset, collection methodology, data availability, and discusses potential challenge problems being developed using this data. Datasets and future challenge announcements will be available via https://dcc.mit.edu.

## I. Introduction

High Performance Computing (HPC) centers have traditionally focused on large scale simulation applications developed in C and Fortran. In recent years, these HPC centers are experiencing a significant diversification of workloads including a transition towards high level programming languages such as Python, Julia, and MATLAB for a wide variety of scientific applications, as well as a significant number of workloads involving Artificial Intelligence (AI) and Machine Learning (ML) [1]–[3]. As AI/ML applications have become an increasingly larger component of the workload, HPC centers have had to evolve in order to find a balance in the continuum between high performance and high productivity [4]–[7]. This evolution requires the right combination of hardware platforms as well as performance and productivity tools, which have become more critical and complex at the same time. Additionally, the mere availability of high-performance accelerators such as GPUs and frameworks such as TensorFlow and PyTorch are not in and of themselves a guarantee of optimal resource usage. A significant amount of tuning of optimized libraries such as cuDNN, NCCL, CUDA-aware MPI, along with appropriate process launch mechanisms and framework configurations, are required to achieve near-peak performance. Thus, in order to optimize the use of expensive, finite, high-performance hardware, it is imperative to collect, integrate, fuse, and analyze the data of various system components such as storage, hardware, networking, applications, power, and other sensors from a cluster/datacenter, with the goal of ensuring high performance, high availability, efficient allocation, and appropriate use of resources. Additionally, the analysis of data from various compute workloads in a cluster has the potential to inform pathways to the optimization of cluster operations as well as user code.

Along with the significant increase in AI/ML compute workloads, the number and size of datasets that accompany these analyses have also grown significantly. This compounds the challenge of developing efficient data input pipelines for training and inference operations that do not adversely affect the application performance. Thus, there is an increasing need to have more efficient data storage and better understanding of bottlenecks within the cluster/datacenter architecture as well as identifying components that are prone to failure or sub-optimal use. These combined effects have driven demand for novel techniques, such as applications of AI/ML, in identifying and addressing these problems and made datacenter usage analysis an active area of research. Despite this, the availability of large scale, open source, labeled datasets that can enable this research is limited, with few datasets available publicly. The dataset we describe in this paper aims to address this gap.

In this paper, we describe the collection of a rich dataset from the MIT Supercloud petascale cluster. The dataset includes parsed logs from the scheduler, file system, compute nodes, CPU and GPU usage time series data, as well as sensor data from physical monitoring of the facility housing the cluster itself. Our goal in creating this dataset is to foster innovations in AI-based approaches to the analysis of datacenter monitoring systems. We anticipate that this rich dataset can help enable goals such as:

- Prediction and early identification of system failures

- Optimization of system scheduling for improved resource consumption
- Identification of optimization pathways for user jobs
- Optimization of datacenter power consumption
- Detection of policy violations

The paper is organized as follows: Section II discusses prior work in the area of datacenter datasets. In Section III and IV we describe the data itself in more detail as well as the data collection and anonymization approaches. Section V provides statistical analysis of user-submitted GPU jobs and GPU utilization. Sections VI and VII provide details on potential challenge problems being developed using this dataset and accessing the data respectively.

## II. Background and Related Work

Several HPC cluster and commercial cloud datasets are already publicly available. These include datasets available on archives such as the Parallel Workloads Archive (PWA) [8], [9], the Grid Workloads Archive (GWA) [10], and the Failure Traces Archive (FTA) [11]. Additional datasets include the Google Cluster-Usage Traces [12]–[14], the Atlas Cluster Trace Repository [15], the Philly-Traces [16] dataset from Microsoft, and the Blue Waters System Monitoring Data Set [17]. These four datasets, along with our MIT Supercloud Dataset, are summarized in Table I. The remainder of this section summarizes the above archives and then discusses these four datasets in detail. Section III provides details on the MIT Supercloud Dataset.

The PWA is a collection of raw workload/scheduler logs collected from parallel machines across the world. In addition to raw logs, the PWA hosts workload models and bibliographic references to papers related to workload issues. At time of access, the PWA hosts forty workload logs collected between 1993-2018, with the most recent addition in September 2019. Among the papers listed on the PWA, analyses consists of regression applied to scheduler optimization [18] and host overload detection [19], the application of clustering techniques to learning workload characteristics [20] and grouping web services into different classes [21], and the prediction of CPU, memory, and I/O usage to optimally align scheduler jobs with available resources with a focus on reducing energy consumption [22]. Additionally, two papers listed on the PWA utilize the Google Cluster-Usage dataset mentioned above to compare and contrast cloud versus grid work load and host load characteristics [23] and perform statistical analysis of task and resource allocation related to Google applications [24]. The PWA website also includes links to other grid workload trace archives such as GWA [10] from Delft University of Technology and the FTA [11]. At time of access the GWA listed thirteen datasets, of which nine were available for download and the FTA listed twenty seven available datasets. Although the number and variety of datasets found on the above mentioned archives (as well as other listed repositories on the PWA) are too numerous to describe in detail, to the best of our knowledge, the common feature among them is that these datasets are primarily limited to workload manager logs.

Google published versions 1 and 2 of the Google Cluster-Usage in 2010 [12] and 2011 [13], respectively, with version 3 being published in 2019 [14]. The goal of this dataset was to make publicly available the complexities of Google's HPC scheduling workload and facilitate research in the area of HPC cluster management. In addition to a variety of scheduler log data, the 2011 release included summary statistics of CPU time series data sampled at one second intervals over a five minute window. However, the actual time series data was not included. In 2019, CPU usage histograms for the five minute windows were added, along with information on alloc sets (groups of resources reserved for a job [25]) and job-parent information for jobs such as MapReduce. For further details, we refer readers to the original release documentation.

The Atlas Cluster Trace Repository [15] is a repository of four datasets: two collected from Los Alamos National Laboratory (LANL) and two collected from Two Sigma, a private hedge fund. The goal of this dataset was to reduce the reliance on Google's Cluster-Usage Traces dataset, which the authors of [26] claim is leading to overfitting in the context of HPC cluster management research. In order to produce a more diversified HPC dataset, the Atlas repository includes traces that represent a variety of workloads ranging from web services to high performance computing with median job runtimes up to 4-5 times longer than those in the Google dataset. However, to the best of our knowledge, the Atlas repository does not include time series data related to CPU/GPU utilization.

The Philly-Traces [16] dataset includes deep neural network (DNN) training workloads on Microsoft's internal clusters, with the goal of studying cluster utilization for DNN workloads in a multi-tenant environment. In particular, the focus of this dataset and associated research is on scheduling, constraints, proximity of allocated resources, and failures in the context of GPU allocation for DNN workloads. The dataset is drawn from three primary sources: scheduler logs, standard output and error logs, and per-minute statistics on CPU, GPU, memory, and network utilization. Job runtimes range from minutes to hours and even weeks.

Perhaps the most comprehensive dataset available is the Blue Waters System Monitoring Data Set [17] which includes I/O data, statistics on node usage, memory allocations, CPU or GPU performance (sampled at 60s), file transfers, data calls, communication link status, and many more detailed metrics including hardware counters.

## III. MIT Supercloud Dataset description and collection methodology

The data being made available as part of this release were collected from the MIT Supercloud TX-Gaia cluster [27]. The TX-Gaia system is a heterogenous cluster consisting of a set of GPU-accelerated nodes and another set of CPU-only nodes. The first partition has 224 nodes with two 20-core Intel Xeon Gold 6248 processors with a total 384GB of RAM and two NVIDIA Volta V100 GPUs with 32GB of RAM each.

| Dataset | Collection Time Frame | Job Count | Scheduling Data | Compute Characteristics |
| --- | --- | --- | --- | --- |
| Google Cluster Usage Traces v3 (2019) | 1 Month | 672,074 (2011) | Machine ID, capacity, attributes, type of event and resource requests | Sample CPU at 1s, CPU Summary statistics over 300s interval, not actual time series data |
| Atlas Cluster Trace Repository | 3-60 Months | 96,260 | Start/stop times, time budgets, job exit status and properties | None |
| Philly Traces | 2.5 Months | 117,325 | Start/stop times, GPU requests and allocations, Stdout and stderr logs | CPU/GPU at 60s, CPU, memory, network, and GPU utilizations |
| Blue Waters System Monitoring Data Set | 80 months | 10.5 Million | I/O data, statistics on node usage, memory allocations | CPU/GPU at 60s. CPU/GPU memory utilization, CPU load averages, GPU temperature and power |
| MIT Supercloud Dataset | 6 Months (ongoing) | > 1 Million | I/O data, statistics on node usage, memory allocations, hardware utilization, physical infrastructure | CPU time series at 10 sec., GPU time series at 100ms, load, user and process count, file system latency on compute nodes sampled every 5 minutes |

TABLE I
HPC CLUSTER AND DATACENTER/CLOUD DATASETS

| Measured Quantity | Unit/Indicator | Measured Quantity | Unit/Indicator |
| --- | --- | --- | --- |
| Total Power | KWH | Fire/Smoke | Boolean |
| Line voltage, current | Volts, Amps. | Water detection | Boolean |
| Outside air temperature | Celcius | Humidifier | Boolean |
| Outside air humidity | % humidity | Heater over temperature | Boolean |
| Aisle temperature, pressure, humidity | Celcius, Bar | Exhaust fan | % speed |
| Cold/Hot aisle pressure differential | Bar | Cold aisle door status | Boolean |
| Pressure sensor failures | Boolean | Cold and hot aisle temperature | Celcius |
| Power supply alarm | Boolean | Cold and hot aisle pressure | Bar |
| Direct Expansion (DX) Cooling Enable | Boolean | Heater Enable | Boolean |

TABLE II
PHYSICAL INFRASTRUCTURE MONITORING OVERVIEW

The second partition is a CPU-only partition and consists of 480 nodes with two 24-core Intel Xeon Platinum 8260 processors each with 192GB of RAM. The central file system is a Lustre high performance parallel file system running on a 3-petabyte Cray L300 parallel storage array, consisting of one metadata server and four data servers. All of the login nodes, compute nodes and service nodes (scheduler, file system metadata server, file system data servers, etc.) are interconnected through an Arista Ethernet Core Switch with all links being 25-Gigabit Ethernet connections. The exceptions are the file system data servers which are connected to the Arista Core Switch with four channel-bonded 100-Gigabit Ethernet connections for each of the four data servers. This core switch network is a one-to-one subscription, non-blocking star network (no leaf-spine topology), in which every computer is equidistant from one another. The data collected includes physical sensor data, scheduler logs, node level data, time series from jobs, and file system data.

*Physical monitoring data:* The cluster is housed inside of a purpose-built modular Performance Optimized Datacenter (EcoPOD 240a), manufactured by Hewlett Packard Enterprise. The EcoPOD houses 44 racks of IT equipment in two rows of twenty two racks (rows A and B) and leverages hundreds of sensors to dynamically maintain optimal operating conditions. The EcoPOD's rich data source includes power data for IT and heating, ventilation, and air conditioning (HVAC), exhaust fan speeds, and current PUE. Detailed environmental data is available across twelve collection points corresponding to the Direct Exchange (DX) air conditioning units on both IT rack rows. Monitoring includes the temperature, humidity, fan speeds, and damper positions providing details of the operating environment at almost rack level granularity. A similar suite of data is available for the outside environment which can provide insight into the optimal conditions to take advantage of economizer operating modes. We expect that analyses of this data will provide insight into the effect of submitted jobs on the datacenter environment control and potential strategies to optimize job submission with the goal reducing the carbon footprint. Table II lists the major categories of data measured by the sensors and corresponding units as appropriate.

*Scheduler data:* MIT Supercloud uses Slurm – the simple

| Metric | Description | Metric | Description |
|---|---|---|---|
| *id_job | Unique job ID | cpus_req | Number of CPUs/cores requested |
| *id_array_job | Unique array job ID | derived_ec | Derived exit code |
| id_array_task | Unique array job ID | derived_es | Derived exit state |
| *id_user | Unique User ID | exit_code | Exit code |
| *kill_requid | User ID of the user who ended the job | array_max_tasks | Maximum number of tasks per array |
| nodes_alloc | Number of nodes allocated to job | array_task_pending | Number of pending array tasks |
| *nodelist | List of nodes on which job executed | constraints | Type of processor on which execute job |
| flags | Additional Slurm flags specified | time_start | Date/time job started |
| mem_req | Amount of memory requested | time_end | Date/time job ended |
| partition | Queue/partition of job | time_suspended | Date/time job was suspended |
| priority | Job priority when job started executing | track_steps | Indicates whether job steps are tracked |
| state | Final job state (completed, failed, etc.) | tres_alloc | Trackable resources allocated |
| timelimit | Time limit for job | tres_req | Trackable resources requested |
| time_submit | Date/time when job was submitted | job_type | Text description of job type |
| time_eligible | Time job is eligible to run | gres_used | Number of GPUs used |

TABLE III
SLURM LOG DATA FIELDS

linux utility for resource management – for resource management and job scheduling [28]. Slurm includes a database which provides details for every jobs that ran on the system over the collection period. These job details include the job name, resources requested/allocated, work directory, job run times, and many more. We make a subset of the fields in the Slurm database available as part of this dataset. Some fields are removed to ensure privacy, and other fields, such as the job ID, are anonymized (as described in Section IV). Additionally, fields that have been deemed to not provide useful data for analysis are removed. Table III shows the fields that are present in the released data. Entries marked with an asterisk indicate anonymized values.

*Slurm time series data:* The Slurm scheduler has the ability to capture time-series data for all jobs running on the system. This capability can be used to collect usage statistics at a specified interval from the CPUs, node memory, file I/O activity, and power usage. We use the `acct_gather_profile/hdf5` plugin to collect detailed data on all jobs running on the cluster. The data is generated by sampling various performance data either collected by Slurm, the operating system, or component software. The plugin records the data as a time series and also accumulates totals for each statistic for the job [29]. We collect this data at a ten second interval. Data collection is started for each job as part of the job prolog and stopped in the job epilog. The data is written to the local disk of the job's compute node(s) as the data is collected and copied to central storage once the job has ended. Collecting this data on the local disk ensures transparent operation and minimizes any impact on the shared file system or on users' jobs. Table IV shows the data fields monitored using this approach. Along with this time series data, we also generate a time series summary which includes the minimum, maximum, and average of the fields listed in this table for each job.

*GPU time series data:* Along with the CPU usage, we also monitor all GPUs assigned to a job using the `nvidia-smi` utility [30]. Using this tool, we can monitor the GPU usage on a node and collect metrics on SM (Streaming Multi-processor) utilization, GPU memory footprint, power draw, and the PCIe bandwidth. This data is collected on every node and every GPU assigned to a job. The data is collected every 100ms and written to the local disk on each node, similar to the Slurm time series data, and moved to central storage at the end of the job. Collection is started and stopped automatically using the Slum prolog and epilog, respectively. Table V shows the tracked fields on the GPU.

In the context of GPU time series data, a subset of the MIT Supercloud Dataset contains labeled DNN jobs. Some of the representative DNN model architectures included in this subset are VGG-11/16/19, Inception-3/4, Resnet-50/101/200, and several U-Net variants. In the future, this will be augmented with jobs that run training and inference on Natural Language Processing (NLP) models, transformers, and non-DNN machine learning approaches.

*Node data:* Node level data is important for monitoring the health of a compute node and identifying failures from jobs running on the node. The nodes may be scheduled for exclusive use by a job or may share resources with several jobs simultaneously. This data provides an alternate view of compute node utilization and is collected across the entire cluster independent of the scheduler and jobs running on the system. The data is collected every five minutes on each node. The fields that are monitored and collected include the user process counts, system load, CPU memory usage, number of Lustre remote procedure calls (RPC) made on the node in the time interval, and the current latency of the file system. Taken together, the above metrics provide insight into the health of a given compute node and can potentially be used to developed baselines for "normal" behavior of the system at the node level. A complete listing of the above node level metrics is given in Table VI.

*Lustre file system logs:* MIT Supercloud uses the Lustre parallel distributed file system. This dataset will also include Lustre metadata server logs, which provide important information about cluster-wide I/O char-acteristics. The Lustre metadata logs include the following low-level file system operations: `open`, `close`, `link`, `unlink`, `mknod`, `mkdir`, `rmdir`, `rename`, `getattr`, `setattr`, `getxattr`, `setxattr`, `statfs`, `sync`, `samedir_rename`, and `crossdir_rename`. The Lustre file system logs will be parsed and made available as csv files, similar to the rest of the dataset.

| Metric | Description |
| --- | --- |
| *id_job | Unique job ID |
| *Node | Node name(s) on which job executed |
| Step | Slurm job step |
| ElapsedTime | Sample time from start time of 0 |
| CPUFrequency | CPU clock frequency |
| CPUTime | Time spent on compute by CPU |
| CPUUtilization | CPU utilization by job |
| RSS | Resident Memory Footprint Set Size |
| VMSize | Virtual memory used by process |
| Pages | Linux memory pages |
| ReadMB,WriteMB | Amount of data read/written |

TABLE IV
SLURM TIME SERIES DATA FIELDS

| Metric | Description |
| --- | --- |
| *id_job | Unique job ID |
| *Node | Node ID |
| timestamp | Timestamp of time series sample |
| gpu_index | GPU Index ID for given Node |
| utilization_gpu_pct | Percentage of GPU utilized |
| utilization_memory_pct | Percentage of memory utilized |
| memory_{free,used}_MiB | Available and used GPU memory |
| temperature_gpu,memory | GPU temperature |
| temperature_memory | GPU Memory temperature |
| power_draw_W | Power drawn |
| pcie_link_width_current | PCI express bus width |

TABLE V
NVIDIA TIME SERIES DATA FIELDS

| Metric | Description |
| --- | --- |
| *Node | Node ID |
| *UserPIDCount | User Id and user process count |
| Time | Timestamp when data was collected |
| FSlatency | Latency of file system access |
| LoadAvg | Average load on the node |
| MemoryFreeInactiveKB | Memory usage on node |
| LustreRPCTotals | Count of Lustre remote procedure calls |

TABLE VI
NODE LEVEL DATA FIELDS

At the time of writing, we have collected CPU time series data and job statistics from over one million jobs and almost 90,000 time series traces from GPU usage. This data collection is ongoing and we expect to augment the data with additional fields/metrics and/or add additional data over time.

## IV. DATA PRE-PROCESSING

As mentioned earlier, one of the primary uses of the MIT Supercloud dataset is performing AI/ML analysis on data collected from an HPC environment. However, the monitoring data and logs described in the previous sections are not AI-ready. Several pre-processing steps had to be implemented in order to make the data more amenable for use in developing AI/ML algorithms. The pre-processing pipeline includes removing outliers and incomplete data, parsing log files, adding missing metadata, and converting the data to a format that is easily usable in commonly used programming languages. In addition to this, Slurm time series data must be extracted from HDF5 files and appropriate metadata needs to be added to link the time series with the corresponding job. Similarly, the GPU monitoring data also needs the addition of the corresponding jobid and node name in order to link each dataset with the corresponding job. The `nvidia-smi` tool used to collect GPU usage data is not scheduler-aware and does not automatically associate the monitored time series with the Slurm job ID. Further, since the compute nodes have two GPUs each, there are instances where each GPU on a node may be assigned to a different job/user, which requires disambiguation for analysis. Thus, additional metadata such as the job ID and node name was added to the GPU time series data to link each GPU trace to the corresponding job and the specific node on which the data was collected.

One important aspect of pre-processing this data is the anonymization of the data in order to preserve privacy. The Slurm logs contain identifiable information such as the user ID, job name, job work directory, and node name. There are several approaches to achieve the goal of anonymization. Mature encryption methods such as RSA offer strong guarantees; however, these may lead to a loss in data linkages across different collection tools. That is, values that are similar before the hashing operation will no longer be similar after, which leads to a loss of information and potentially makes the analysis more difficult. As a first step, we completely remove information that is not relevant to the envisioned AI/ML analysis. Thus, the job names and job work directories are not included in the dataset. Removal of these fields from the scheduler logs also ensures that user privacy is preserved. Similarly, MAC addresses of nodes, Linux kernel versions, and other site-specific constants are also removed because they do not provide any information relevant to hardware usage. However, job IDs, user IDs, and node names are key identifiers that are essential to the analysis of the data and were anonymized.

The anonymization process encrypted the above identifiers using a SHA-256 hash and salted them with a constant string. We do not intend to publish details of this process to ensure anonymity of the data. Unique job IDs, usernames, and node names were collected from the entire dataset prior to anonymization. Post-anonymization, the original values in all data files were replaced with the new hashed values. This ensured that a job ID in the Slurm dataset had corresponding entries in the CPU and GPU time series data and the data linkage was preserved. We also anonymized all node names occurring in the dataset. The hostnames of the compute nodes in the cluster are of the form A-B-C-D or for special nodes such as the login nodes, the hostname is of a simpler format,

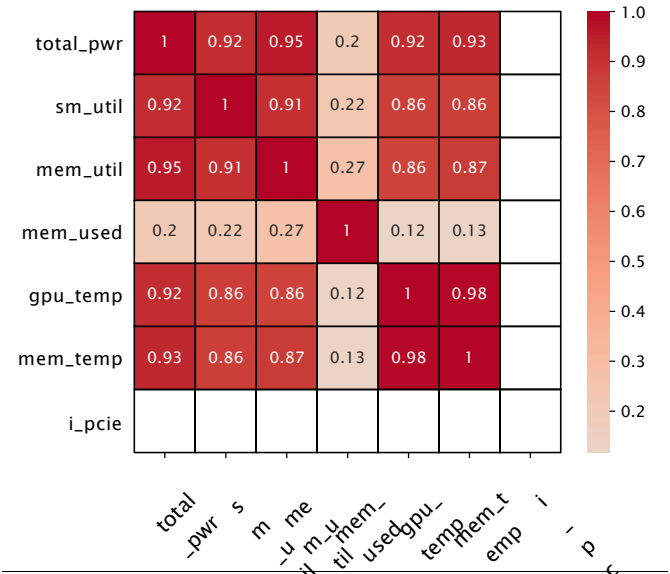

Fig. 2. Spearman correlation of nvidia-smi GPU measurements for all jobs. The measured fields are GPU power (W), SM utilization (%), GPU memory utilization (%), GPU memory used (MB), GPU temperature (°C), GPU memory temperature (°C), PCIe link width current

namely, AA-BB. We treat the first half of the node name as the rack identifier and the second half as the node number in the rack. In all cases, these sets of strings representing hostnames were hashed to get a unique id value for the rack and node identifier and used to create a new identifier in the form `r[]-n[]`. For example, the hostname `d-10-1-1` is anonymized as `r62817-n20038`.

## V. Preliminary analysis

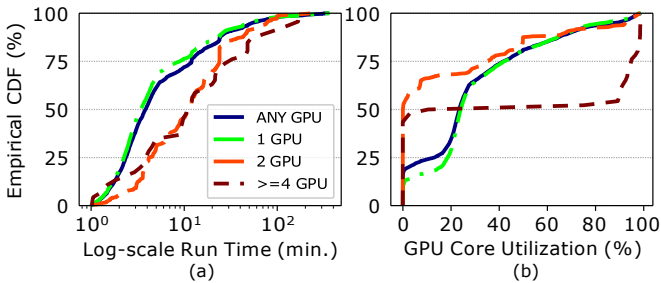

Fig. 1. (a) Empirical CDF distribution of job run time for different number of GPUs used. (b) Empirical CDF distribution for job GPU streaming multiprocessor utilization for different number of GPUs used.

As the MIT Supercloud features GPU accelerations, we are interested in the characteristics of user-submitted GPU-accelerated jobs and how such jobs utilize the GPUs. We break down the job execution time and GPU core utilization by the number of GPUs a job uses, as a job may use multiple GPUs for tasks such as distributed training [31]. For the job execution time empirical CDF shown in Fig. 1(a), the execution time varies over a wide range, from a few minutes to hundreds of minutes. The blue solid curve represents all jobs using any number of GPUs and the median job execution time is around four minutes. Jobs that use a single GPU have a similar execution time CDF as all jobs. This is explained by the fact 85% of the jobs use only one GPU. Jobs that use multiple GPUs have longer execution times than single-GPU jobs. Note that in the MIT Supercloud, every node has two GPUs and 2-GPU jobs can be distributed over a single node, while the jobs that use four GPUs or more are inter-node distribution jobs. The intra- and inter-node distributed jobs have a median execution time of ten minutes.

The GPU SM core utilization also shows different characteristics between jobs using a different number of GPUs, as demonstrated in Fig. 1(b). The single-GPU jobs have a median utilization of 23%. The 2-GPU jobs have a surprisingly lower utilization than the single-GPU jobs, as 50% of such jobs have a utilization close to zero. This is possible when a user requests and is allocated an exclusive node for development, debugging, or exploratory purposes, and the GPUs are not utilized during the majority of these job types. The inter-node distributed jobs demonstrate very distinctive behavior, as half of the jobs have utilization close to zero, which, similar to above, are possibly development, debugging, or exploratory jobs. The other half have very high utilization ($> 80\%$), which are most likely mature jobs.

The GPU usage statistics collected by the `nvidia-smi` tool (Sec. III) show strong correlation between most metrics across jobs. For each job, the GPU time-series measurement is aggregated into average values, and the Spearman correlation coefficient is calculated for all jobs, as shown in Fig. 2. The Spearman correlation is a ranking correlation, the correlation coefficient is between -1 and 1, with 0 indicating no correlation, values close to 1 indicating a high ranking correlation and values close to -1 indicating a high reverse ranking correlation. Linear correlation (or Pearson correlation) is also calculated, but is not as strong as the ranking correlation. From Fig. 2, the power consumption, SM utilization, memory utilization, GPU temperature, and memory temperature are highly correlated, indicating that jobs with a high value of any of these metrics (e.g. high GPU SM utilization) are also likely to have high measurements of the other metrics (e.g. GPU power consumption). The GPU memory used and PCIe link width current do not show much correlation with the other metrics. In fact, the PCIe link width remains consistent for all jobs.

## VI. Potential Challenge Problems

Along with the MIT Supercloud Dataset, we also plan to release a set of challenge problems and associated baseline implementations. Challenge problems are expected to address one or more of the following potential analysis:

- Can AI models better predict job run-times? In most HPC centers, users are required to specify the amount of wall time their job is expected to consume. Typically, jobs get killed when the requested time allocation is exceeded. This can lead to users over-estimating time requirements, which can make scheduling harder and potentially lead to longer wait times. Similarly, in a cloud-computing scenario, better estimates of job run times can help organizations budget appropriately. The ability to more accurately predict job run times may make it possible to better utilize resources through improved scheduling.

- Can time series data be used to develop AI-based tools to enable early prediction of job failure? Jobs can fail for a variety of reasons ranging from hardware failures to excessive resource consumption or bugs in the code itself. In some cases, it may be possible to identify signals in the time series data that can help mitigate certain classes of failures.
- Can we identify similar workloads from anonymized data? Can workloads be identified as AI/ML jobs vs. other kinds of compute?
- Can we establish a baseline for workloads to be able to suggest pathways for optimization or scheduling on hardware resources better suited for the job? Workload characterization may also be beneficial in identifying policy violations (for example, whether a job is mining crypto-currency or running password cracking programs).
- Can we build intelligent systems that can link data from scheduler logs, compute nodes, Lustre logs, and job IDs to provide an easy way to correlate cause and effect? For example, if a node crashes, can we identify which job was most likely responsible (since multiple jobs can share a node) and learn the resource usage characteristics of the identified job to provide better monitoring?
- Given time series data from nodes and the Lustre file system, can we determine what normal usage of the system look like? By modeling the distribution of file system operations such as `getattr`, `setattr`, `mkdir`, `rmdir`, and others, can we develop AI that can identify patterns that indicate excessive load on the file system before it becomes an issue? While it is possible to use simple thresholds based on heuristics, an early indicator of a badly behaved job could potentially mitigate system-wide outages due to a handful of jobs/users. For reference, our team will provide participants with a set of thresholds that are currently used to activate alerts on our monitoring system.

## VII. DATA AVAILABILITY

The MIT Supercloud Dataset, challenge specifications, and links to baseline implementations will be made available via https://dcc.mit.edu. We plan to release the following sets:

- **Supercloud Labeled:** This stand-alone dataset consists of monitoring data from a variety of known AI/ML workloads that use standard, publicly available implementations and datasets. It is anticipated that this subset will be suitable for use as ground truth for characterizing compute workloads. At present, this dataset contains 6,000 jobs and we anticipate augmenting this in terms of number of jobs and AI/ML workload varieties.
- **Supercloud Unlabeled:** This is the complete releasable dataset and currently contains traces from over 500,000 jobs. The dataset will continue to grow, as the collection is ongoing.

In all cases, researchers will be required to accept a data use agreement prior to data download, details of which will be posted on the Challenge website listed above.

## VIII. SUMMARY AND DISCUSSION

Data from the MIT Supercloud consists of complex and diverse data from schedulers, heterogenous hardware, and building management systems. These data can provide invaluable insight into the working of a datacenter to answer questions such as: "Is hardware failure likely?" or "Is there anomolous behaviour in my network?" However, performing cross-data analysis can be challenging due to different sources, data types, and data acquisition rates. For example, some sensors provide rich, time-series datasets that must be aggregated while others provide snapshots of the system that present a static view of the system. Machine Learning and Artificial Intelligence have the potential to offer new insights into many aspects of the datacenter such as system and workflow characterization, predictive modeling, smart scheduling, resource optimization, and several others.

In this paper, we present a comprehensive dataset from the MIT Supercloud cluster that includes system logs such as file system monitoring, node level logs which include resource utilization (CPU, Memory), both CPU and GPU time series data from a subset of all jobs running on the system, as well as scheduler logs. The data is presented in an easily readable CSV format. All data is anonymized to preserve user privacy while retaining relationships between jobs and resource usage. We also present potential challenge problems that can be developed using this dataset and will be announced separately.

ACKNOWLEDGMENTS

The authors acknowledge the MIT Lincoln Laboratory Supercomputing Center (LLSC) for providing HPC resources that have contributed to the research results reported in this paper. The authors wish to acknowledge the following individuals for their contributions and support: Bob Bond, Tucker Hamilton, Jeff Gottschalk, Tim Kraska, Andrew Morris, Charles Leiserson, Dave Martinez, John Radovan, Steve Rejto, Daniela Rus, Marc Zissman.